\documentclass[aps,prl,reprint,groupedaddress]{revtex4-1}
\usepackage{amsmath}
\usepackage{amssymb}
\usepackage{graphicx}

\begin{document}
\title{Inferring the Rate-Length Law of Protein Folding}

\author{Thomas J. Lane}
\author{Vijay S. Pande}
\email[]{pande@stanford.edu}
\affiliation{Department of Chemistry, Stanford University}

\date{\today}

\begin{abstract}
We investigate the rate-length scaling law of protein folding, a key undetermined scaling law in the analytical theory of protein folding. We demonstrate that chain length is a dominant factor determining folding times, and that the unambiguous determination of the way chain length correlates with folding times could provide key mechanistic insight into the folding process. Four specific proposed laws (power law, exponential, and two stretched exponentials) are tested against one another, and it is found that the power law best explains the data. At the same time, the fit power law results in rates that are very fast, nearly unreasonably so in a biological context. We show that any of the proposed forms are viable, conclude that more data is necessary to unequivocally infer the rate-length law, and that such data could be obtained through a small number of protein folding experiments on large protein domains.
\end{abstract}

\pacs{}

\maketitle

\section{Introduction}

A deep understanding of protein folding must involve a description of the general mechanisms involved. It is reasonable to suspect this will consist of a simple model, based on microscopic physics, expressed in a simple mathematical language. Such a model would show how biological sequences are able to employ physics to spontaneously self-assemble into intricate molecular machines. 

Simple models of this sort often start by postulating a mechanism of folding, and then derive the consequences of that mechanism \cite{Lane:2012ba, Gutin:1996iu, Cieplak:2003cq, Cieplak:1999wa, Zwanzig:1992be, Zwanzig:1995wi, Thirumalai:1995tn, Li:2002iv, Naganathan:2005dr, Koga:2001ii, Finkelstein:2007vi, Wolynes:1997te}. This suggests it might be possible to infer the general mechanisms of protein folding by verifying the specific predictions of these models. For a simple model of protein folding, however, there is a limited set of \emph{general} experimental trends that can be readily predicted. One such experimental trend is the law governing how folding times scale with chain length. This is perhaps the simplest comparison of theory and experiment possible, but has not yet been unambiguously inferred despite nearly two decades of active research \cite{Thirumalai:1995tn, Gutin:1996iu}.

The rate-length law is also an interesting result in and of itself. Such a law can be viewed as a statement of the computational complexity of protein folding -- given a problem of size $N$ (residues), how does one expect the time-to-solution (folding) to scale? Levinthal pointed out that an exhaustive search would result in exponential scaling, and suggested that this would result in unreasonably large folding times \cite{Levinthal:1969uy}. Thus, in many ways, a resolution to Levinthal's paradox is likely to be phrased directly as a rate scaling law, either non-exponential (polynomial) or exponential with an explicitly small exponential factor.

A number of issues complicate inferring such a law from experiment, most importantly the fact that available kinetic data on protein folding spans a very limited range of chain lengths - about 30 to 300 residues \cite{Bogatyreva:2009jz}. The statistical power of the data is inherently limited by the fact that protein domain sizes barely span a single order of magnitude, and that most studies of folding have focused on small, well-behaved model systems.

Further complicating the inference of the rate-length scaling law is the fact that chain length is certainly not the only factor affecting folding rates. In fact, it has been argued that it is a fairly weak predictor of the folding time \cite{Plaxco:1998ff, Nakamura:2005iu}. For instance, there seems to be some correlation of folding times with topological complexity of the native state, such that if two proteins have the same number of residues, but different folds, they may take different amounts of time to fold \cite{Plaxco:1998ff, Ivankov:2003dw}. Moreover, even protein mutants with the same native structure can have at least 3 orders of magnitude variation in their folding rates \cite{Lawrence:2010bs}. Thus, we expect that experimental data on the scaling of folding time with chain length should be very noisy, and difficult to statistically estimate.

Nonetheless, there does seem to be a significant correlation between the number of residues ($N$) and folding times ($\tau$). Many different forms of the scaling law have been predicted, but all fall into one of three basic classes. Shaknovich \cite{Gutin:1996iu}, Cieplak \cite{Cieplak:2003cq, Cieplak:1999wa}, and co-workers have proposed a power-law, $\tau \sim N ^ {\nu}$. We recently constructed a model that suggested exponential scaling, $\tau \sim e^{\alpha N}$ \cite{Lane:2012ba}, consistent with predictions made by Zwanzig \cite{Zwanzig:1992be, Zwanzig:1995wi}. Finally, Thirumalai \cite{Thirumalai:1995tn, Li:2002iv}, Mu\~{n}oz \cite{Naganathan:2005dr},  Takada \cite{Koga:2001ii}, Finkelstein \cite{Finkelstein:2007vi}, and co-workers have suggested a stretched exponentials, $\tau \sim e^{ \alpha N^\beta }$, with $\beta$ as $1/2$ or $2/3$. Wolynes has proposed the law may conditionally change between all four suggested models \cite{Wolynes:1997te}.

In what follows, we develop and apply two methods for choosing between these models and evaluate how each proposed model performs.

\section{Modeling}

Below, we outline two complementary methods for inferring which proposed scaling law is the most reasonable. First, we present a method capable of fitting each model's parameters to the known data, and examining how well each model explains the data. Next, we investigate a second discriminatory method, which proposes that folding times must be below a certain threshold value to be biologically viable. It has been demonstrated that in the crowded milieu of the cell, proteins must fold rapidly to avoid aggregation or degradation. We suggest that this implies that we can check the reasonableness of any model by seeing if its prediction for this threshold time is reasonable with what has been observed empirically in biology.

In this study we focus on single-domain globular proteins. Kinetic data for the folding times of proteins were taken from the KineticDB \cite{Bogatyreva:2009jz}, which reports protein folding times at zero denaturant, near room temperature, and under neutral pH. Other data sets exist \cite{Ouyang:2008ch, DeSancho:2011fa, Ivankov:2010hy}, but were not consistent with one another - despite this, they yielded very similar results (see SI, Fig.~2 and Table I).

\subsection{Direct Method: Likelihood Maximization}

We want to estimate the parameters for each proposed form of the scaling law. In what follows, we adopt a model that accounts not only for this scaling law, but all other factors (topology, experimental conditions, etc.) via a random Gaussian component. Thus, by fitting each model, we not only learn parameters for each proposed model, but also get an estimate for the relative importance of these other factors in determining folding times.

We assert the following model for the folding time,
\begin{equation} \label{generic-law}
\log \tau / \tau_0 = f(N) + X
\end{equation}
where
\[
f(N) =
\begin{cases}
\nu \log N & \mbox{power-law} \\
\alpha N & \mbox{exponential} \\
\alpha N^{1/2} & \mbox{stretched exp. (1/2)} \\
\alpha N^{2/3} & \mbox{stretched exp. (2/3)}
\end{cases}
\]
are the proposed folding rate laws, $X$ represents a random variable distributed as a zero-mean Gaussian, $X \sim \mathcal{N}( 0 , \sigma^2 )$, and $\tau_0$ is a fit constant accounting for units of time. By adding $X$ to the logarithm of the folding time (\ref{generic-law}), we model random variation in \emph{relative} terms, and it enters as a multiplicative factor.

Equation (\ref{generic-law}) implies $\tau$ is distributed as a log-normal, with location parameter $f(N)$ and scale parameter $\sigma$. The likelihood of the entire data set (assuming $n$ independent measurements) is
\begin{equation}\label{likelihood}
\mathcal{L} = \prod_{i=1}^n \frac{ 1 }{ \tau_i  \sqrt{ 2 \pi \sigma^2} } \exp \left[ - \frac{ ( \log \tau_i / \tau_0- f(N_i) ) ^2}{2 \sigma^2} \right]
\end{equation}
We have three parameters for each model, $\sigma$, $\tau_0$, and $\alpha$ or $\nu$ for the exponential and power-law families, respectively. 
\begin{figure*}
\includegraphics[width=18cm]{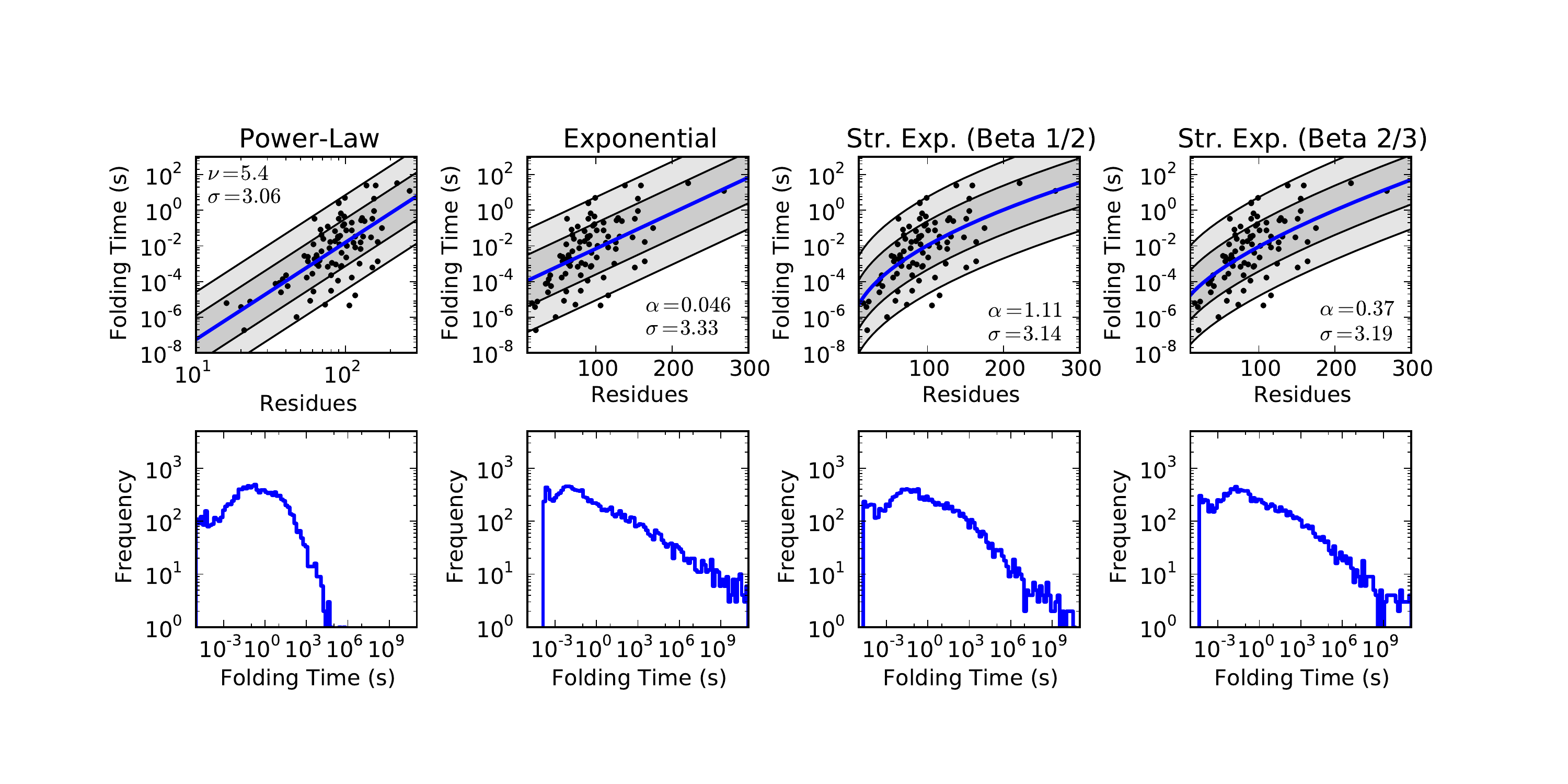}
\caption{The predicted models for the folding rate law, overlaid with measurements of folding times (top), and the putative folding time distributions these models imply (bottom). Parameter values derived from a maximum likelihood fit are displayed, along with intervals indicating the spread in the fit probability distribution. Dark grey shading indicates a factor of $e^\sigma$, while light grey indicates $e^{2 \sigma}$. \label{fits-dists}}
\end{figure*}

We have fit these parameters by maximizing the likelihood $\mathcal{L}$. Model comparison can then be performed by investigating the ratio of the likelihoods of two alternative models (Table \ref{L-ratios}). We have adopted the simple likelihood approach (versus a full-fledged Bayesian analysis) because the number of fit parameters are small and equal for each model, the models are simple and low-dimensional, and we have little prior information about the parameters. See the supplemental information, Fig.~4 and Table II for a Bayesian analysis and comparison.

\subsection{Indirect Method: Biological Limits}

We postulate that there exists a critical time, $\tau_c$, that places a biological upper bound on folding times. Specifically, if a protein folds slower than this time ({\it i.e.} $\tau > \tau_c$) then that protein will be much more likely to aggregate during the course of folding, and therefore is evolutionarily selected against. 

The majority of biologically observed proteins should have folding times less than $\tau_c$, but we postulate that some proteins will have \emph{greater} times. These proteins are those that receive help folding from chaperones or other cellular machinery. It has been estimated that about $C \approx 10\%$ of proteins fall into this category \cite{Hartl:2009bl}.

Together, these assumptions allow us to build a model for the predicted distribution of protein chain lengths. The size distribution of domains (SI Fig.~1) can be roughly approximated by a Gaussian with parameters $\mu_N$ and $\sigma_N$. In that case, 
\[
\int_{f^{-1}(\tau_c / \tau_0)}^{\infty} \frac{1}{\sqrt{ 2 \pi \sigma_N^2} } \exp \left[ \frac{ (N - \mu_N) ^2 }{ 2 \sigma_N^2} \right] = C
\]
where $f^{-1}(\tau_c / \tau_0) = N_c$ is the chain length corresponding to $\tau_c$ for a specific model (power law, exponential, etc.), and $C$ is the percentage of proteins with folding times slower than $\tau_c$. 

This framework is, of course, an approximation. There are undoubtedly many other factors affecting the optimal sizes of proteins beyond merely their folding times. Metabolic efficiency, structural packing constraints \cite{Xu:1998uw, Shen:2005ip}, and the behavior of specific proteins in their local cellular environments certainly play a role. Nonetheless, the concept of an upper limit to the folding times is reasonable, and our aim here is to simply extract some general comments about the reasonableness of predicted folding times, rather than make quantitatively accurate predictions.

\section{Results}
\begin{table} 
\caption{Likelihood Ratios of $\mathcal{L}$-Maximized Models \label{L-ratios}}
\begin{ruledtabular}
\begin{tabular}{l c c c c}
Model \footnotemark[1] & 	Pr. Law &	Exp. &	S. E. 1/2 & 	S. E. 2/3  \\
Power Law     	 &       	   	 	 	&$1.59 \cdot 10^{3}$ 	&$7.98 \cdot 10^{0}$ 	&$3.82 \cdot 10^{1}$  \\
Exponential   	 &$6.30 \cdot 10^{-4}$ 	 &	 	 		  	&$5.03 \cdot 10^{-3}$ 	&$2.41 \cdot 10^{-2}$  \\
S. E. 1/2 	 &$1.25 \cdot 10^{-1}$ 	 &$1.99 \cdot 10^{2}$ 	& 	 	 			&$4.79 \cdot 10^{0}$  \\
S. E. 2/3 	 &$2.61 \cdot 10^{-2}$ 	 &$4.15 \cdot 10^{1}$ 	&$2.09 \cdot 10^{-1}$ 	& 	              		\\
\end{tabular}
\end{ruledtabular}
\footnotetext[1]{Primary model is on the left, alternate model along the top - thus, a larger number favors the model in the leftmost column.}
\end{table}

Direct fitting of all proposed models to the available data yields reasonable results for each (Fig.~\ref{fits-dists}). Each model reports a scale parameter ($\sigma$) of approximately 3, which indicates that 68\% of proteins will have folding times within a factor of $e^\sigma \approx 20$ from the time predicted by the rate law, and 95\% will be within a factor of $e^{2 \sigma} \approx 400$. Since the available data spans folding times of more than 9 orders of magnitude (between $1.9 \cdot 10^{-7}$ and $9.9 \cdot 10^2$ seconds), this demonstrates that chain length captures the majority of variation in folding times, since orthogonal factors (topology, mutations, etc.) account for approximately $400^2 \sim 10^5$ orders of magnitude of variation.

Do the data support any one model? The power-law model is slightly favored by comparing the likelihoods that each model generated the observed data (Table \ref{L-ratios}). In such comparisons, typically a ratio of $10^2$ or greater is considered significant, and often models differ by hundreds of orders of magnitude \cite{KASS:1995vb} -- thus, the power law model is better supported by the data, but only by a modest margin. Further, an attempt to fit the stretched exponential form with $\beta$ as a variable parameter resulted in an unreasonably small value of $\beta$ along with a very large value of $\alpha$, resulting in a fit that is very close to the power law (SI Fig.~3). Finally, the power law model has the smallest fit $\sigma$, indicating that it explains the most variation in the data, and attributes less to orthogonal factors. 

\begin{figure}
\includegraphics[width=8cm]{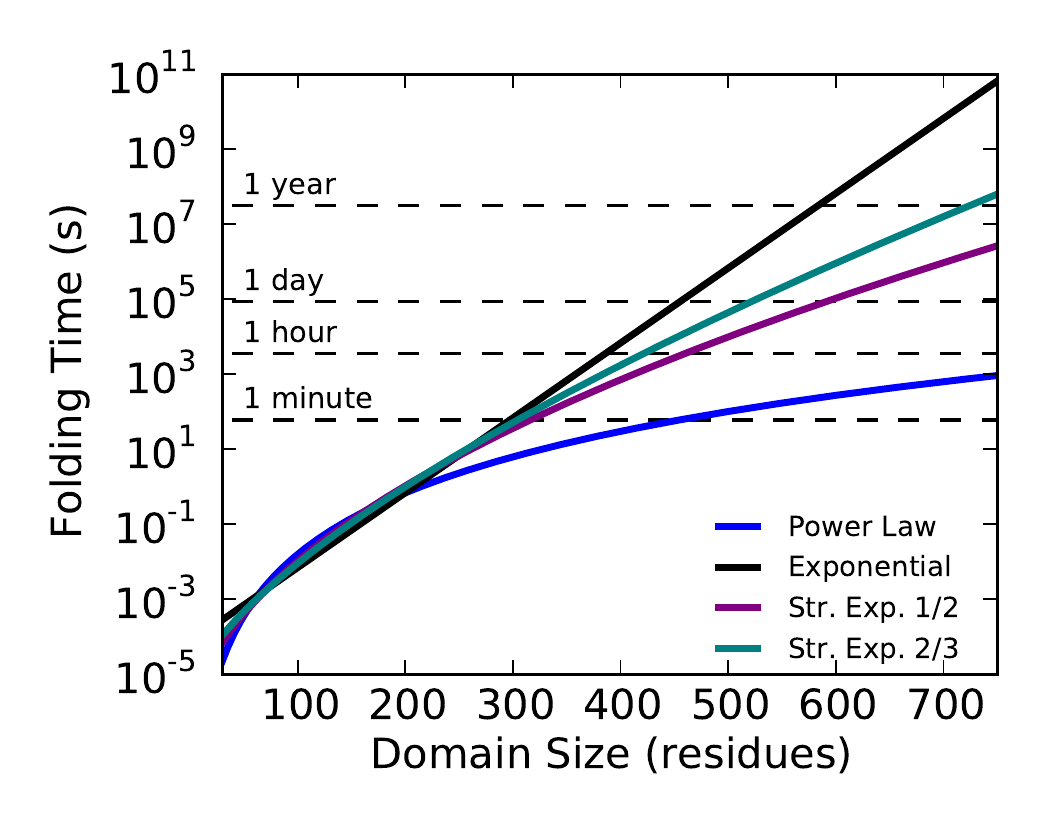}
\caption{The predicted folding times from each model in Figure \ref{fits-dists}, in a direct comparison. Intuitive timescales are denoted for clarity. \label{all-laws}}
\end{figure}

It is clear, however, that there is little difference between the models in the range of available data. These models diverge significantly only for very large proteins (Fig.~\ref{all-laws}).
They yield significantly different predictions for the distribution of folding times, generated by transforming the known distribution of domain sizes into each of the different models (Fig.~\ref{fits-dists}). The most significant differences are in the tails of these distributions, where the exponential forms predict much longer folding times for the largest proteins (Table \ref{time-cdf}).
\begin{table} 
\caption{Estimated Fraction of Protein Domains with Folding Times Greater than Time Indicated \label{time-cdf}}
\begin{ruledtabular}
\begin{tabular}{l c c c c}
			& Hour	 	& Day	 	& Month		&Year \\
Power Law	& 0.41\%	& 0.01\%	& 0.00\%	& 0.00\% \\
Exponential	& 9.56\%	& 5.70\%	& 3.34\%	& 2.46\% \\
S. E. 1/2      	& 5.48\%	& 2.37\%	& 0.95\%	& 0.57\% \\
S. E. 2/3      	& 7.49\%	& 3.53\%	& 1.74\%	& 1.11\% \\
\end{tabular}
\end{ruledtabular}
\end{table}
The power law model predicts no proteins fold in times longer than an hour, while the exponential forms show a significant number of proteins with folding times longer than a day (Fig.~\ref{all-laws}). 

An evaluation of the reasonableness of these folding time distributions is provided by the critical time $\tau_c$ for each model (SI Table III). For reasonable values of $C$, the power law $\tau_c$ is on the order of minutes. The exponential forms, on the other hand, predict $\tau_c$ is on the order of hours. 
\begin{figure}
\includegraphics[width=8cm]{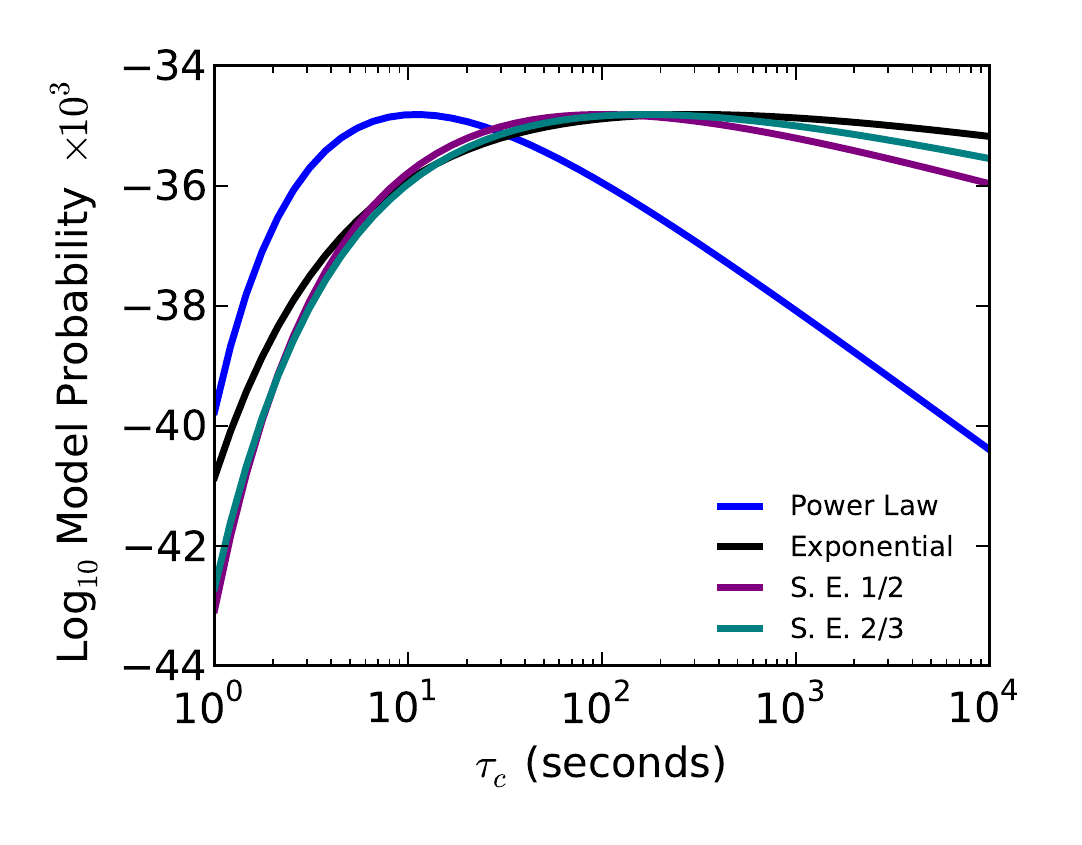}
\caption{The of probability of observing the observed domain sizes (SI Fig.~1) given a rate scaling law and a folding complexity cutoff time, $\tau_c$. Assumes the Gaussian model (with $C=0.10$) described in the main text. Note the y-axis is $\log_{10}$ and divided by $10^3$ for clarity, so small differences on the plot are actually quite large.  \label{tauc-likelihoods}}
\end{figure}
Indeed, picking a reasonable value of $\tau_c$ and calculating the probability of observing the empirically observed domain size distribution (Fig.~\ref{tauc-likelihoods}) shows that for values of $\tau_c \sim 10$ seconds, the power law model is clearly the best. However, for any values of $\tau_c$ greater than $100$ seconds, the exponential laws are much better models.

\section{Conclusions}

Thus, we have a mixed conclusion. While the power law model appears to best explain the available raw data, it results in very fast predicted folding times. The exponential forms, while doing a marginally poorer job of explaining the raw data, yield a distribution of folding times much more in line with what we expect from biology. Given the current available data, no clear victor emerges. 

Previous theories have claimed that simply predicting one of the four laws investigated here is strong evidence in support of that theory. This is manifestly not the case -- not only must the proposed law be reasonable, but it must also predict reasonable parameter estimates, and even then the supporting evidence the rate scaling law can provide given current data is limited. Conversely, the analytical theories mentioned here are not ruled out by the current available data. This is most striking in the case of the exponential form, since exponential scaling of the folding times has often been associated with Levinthal's paradox. This study shows that exponential scaling is reasonable given current experimental data, so long as the exponential scaling constant ($\alpha$) is sufficiently small.

Clear evidence for any one rate law remains missing, however with a few clear examples of very large globular proteins (500 residues or larger) capable of folding unassisted \emph{in vitro}, it might be possible to discriminate between the models proposed here. Figure \ref{all-laws} clearly shows the divergences between predicted folding times for large proteins, and shows how just a few data points in this extreme regime might be able to begin differentiating between the proposed models investigated here.

\bibliography{papers2.bib}

\end{document}


\title{Supplemental Information: Inferring the Rate-Length Law of Protein Folding}

\author{Thomas J. Lane}
\author{Vijay S. Pande}
\email[]{pande@stanford.edu}
\affiliation{Department of Chemistry, Stanford University}

\date{\today}

\pacs{}

\maketitle

\begin{figure}
\includegraphics[width=8cm]{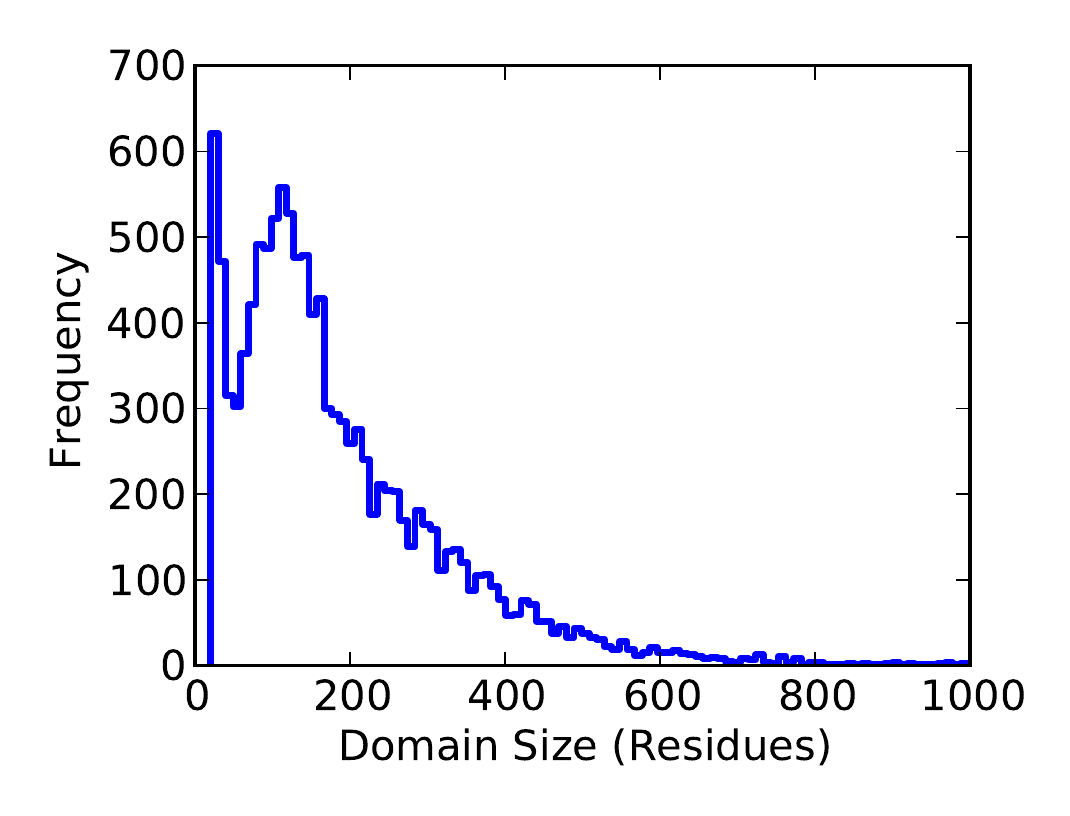}
\caption{The size distribution of non-homologous protein domains listed in the PDB. Of the 12,151 sequences reported, 39 are larger than 1000 residues (0.32\%) - while not shown here for clarity, they were included in subsequent analyses. Data were obtained from the NIH's VAST algorithm ({http://www.ncbi.nlm.nih.gov/Structure/VAST/nrpdb.html}) on 5/5/2012 with a dissimilarity p-value of $10^{-7}$. \label{size-dist}}
\end{figure}

\section{Different Datasets of Protein Folding Kinetic Data}

We are aware of four collections of protein folding kinetic data, here termed ``KineticDB'' \cite{Bogatyreva:2009jz}, ``Liang'' \cite{Ouyang:2008ch}, ``Mu\~{n}oz'' \cite{DeSancho:2011fa}, ``Finkelstein'' \cite{Ivankov:2010hy}; note Finkelstein is reportedly a subset of KineticDB. Each is purported to be extracted directly from the primary literature. Interestingly, while there is much overlap of reported proteins between each dataset, there are systematic inconsistencies between them. Most report that they extrapolate the folding times to zero denaturant - we suspect this is the origin of the discrepancy, but have not undertook a detailed investigation. Instead, we chose the KineticDB dataset because it contained the most proteins, appeared well curated, and restricted entries to proteins at zero denaturant, near room temperature, and at neutral pH.

\begin{figure}
\includegraphics[width=8cm]{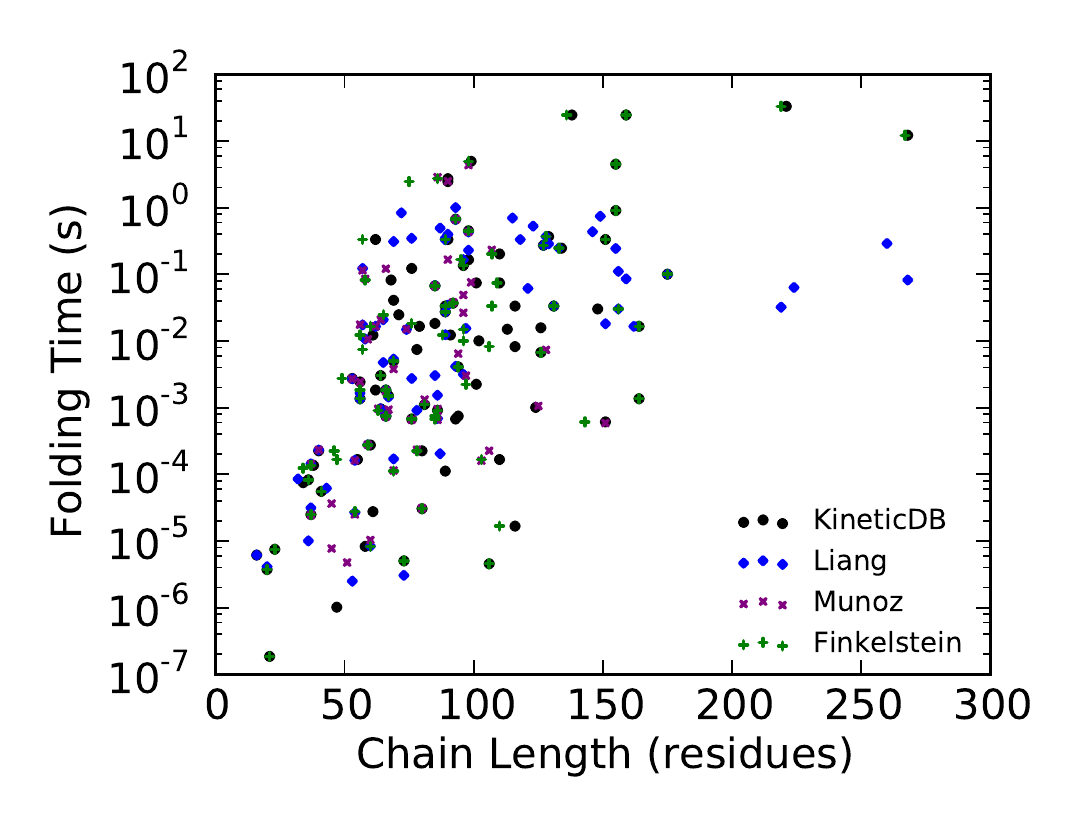}
\caption{The rate-length data from each source used, plotted together. Even though there is some overlap in reported sequences, many identical proteins have different chain lengths or folding times reported. We combined these data for our analysis, eliminating only identical measurements. \label{all-datasets}}
\end{figure}

Interestingly, despite their discrepancies, the datasets appear to be similar on a coarse level (Fig. \ref{all-datasets}). Further, our fitting analysis run on each independently results in parameters that are not too far from one another (Table \ref{dataset-comparison}), and the rest of our results (which primarily follow from these parameters) are very similar between datasets.
\begin{table} 
\caption{Parameters Estimated from Different Datasets\label{dataset-comparison}}
\begin{ruledtabular}
\begin{tabular}{l c c c c c c c c}
 &
\multicolumn{2}{c}{{Pr. Law}} &
\multicolumn{2}{c}{{Expon.}} &
\multicolumn{2}{c}{{S. E. 1/2}} &
\multicolumn{2}{c}{{S. E. 2/3}} \\
Dataset &	$\nu$& 	$\sigma$&	$\alpha$&	 $\sigma$&	$\alpha$&	 $\sigma$&	$\alpha$& 	$\sigma$ \\
\hline
KineticDB		& 	5.44&	3.06&	0.046&	3.32&	1.11&	3.13&	0.37&	3.19\\
KDB (mut) \footnotemark[1] &	4.46&	2.42&	0.040&	2.53&	0.91&	2.45&	0.31&	2.47\\
Liang		&	4.75&	2.54&	0.041&	2.90&	0.94&	2.70&	0.31&	2.77\\
Munoz	 	&	4.79&	3.05&	0.054&	3.14&	1.04&	3.09&	0.37&	3.11\\
Finkelstein	&	5.24&	3.08&	0.048&	3.32&	1.08&	3.17&	0.36&	3.22\\
\end{tabular}
\end{ruledtabular}
\footnotetext[1]{The KineticDB dataset with mutants included. Since there are only a few proteins with mutants, and there are many mutants for these few proteins, this database gives artificially more weight to those individual proteins.}
\end{table}

For this study, we simply extracted all non-mutant entries from the KineticDB and employed their reported chain lengths and folding times in water.

\section{Stretched Exponential with $\beta$ as a Free Parameter}

\begin{figure}
\includegraphics[width=8cm]{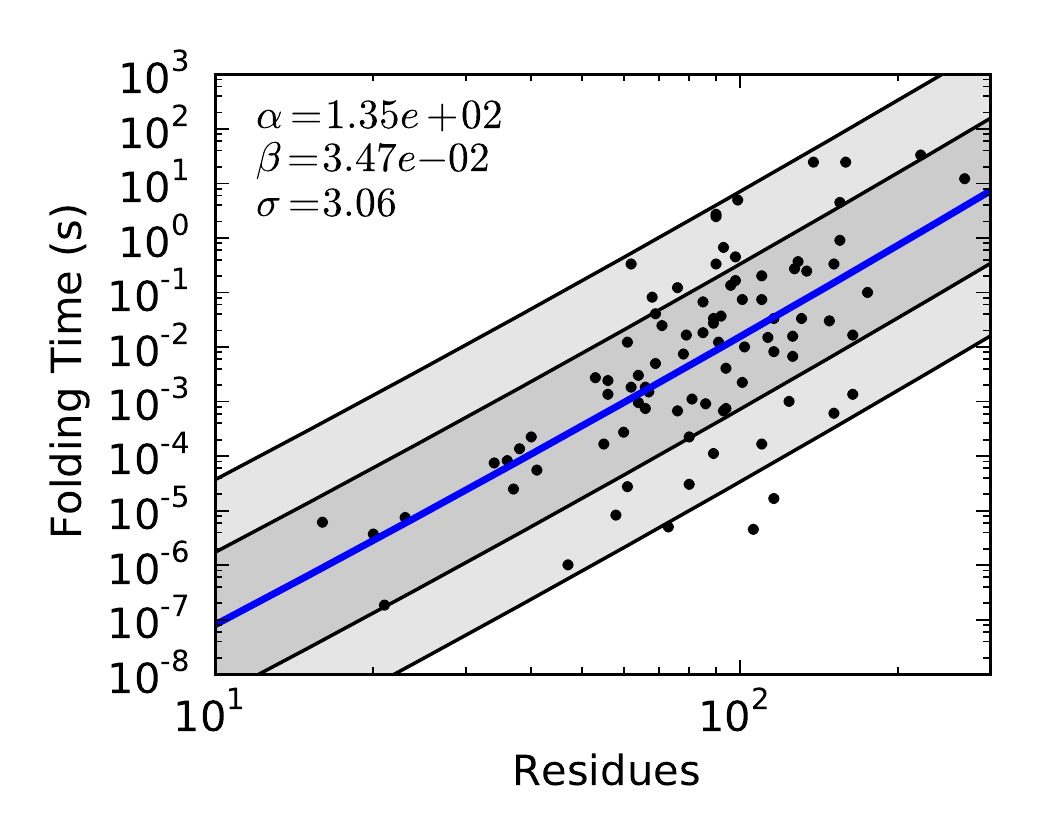}
\caption{The parameter fit for a stretched exponential, $\log \tau \propto \alpha N^{\beta}$, with $\beta$ a free parameter. Notice how the fit is perfectly straight on a log-log plot, a characteristic trait of power laws. \label{floating-SE}}
\end{figure}

In the main text we investigated the specific stretched exponential forms proposed by extant simple models ({\it i.e.} with $\beta$ as 1/2 and 2/3). There is no reason why $\beta$ cannot be simply fit as a free parameter, however, and we have performed this fit (Fig. \ref{floating-SE}).
Remarkably, the most likely parameters are those with $\alpha \approx 10^2$ and $\beta \approx 10^{-2}$, which is far outside the range predicted by theory. With these extreme parameters, the model appears to be a power law form over the range of fit data, and is linear on a log-log plot to about $10^{25}$ residues. This simply verifies that the data are best explained by a power law, and demonstrates that the power law fit with free $\beta$ is sufficiently flexible to capture this fact.

\section{Bayesian Model Comparison}

As mentioned in the main text, it is entirely possible to pursue a full-fledged Bayesian analysis to determine the parameters for each model. Here, we show the commonalities and differences between the maximum likelihood approach pursued in the main text and a Bayesian approach. We show that it makes little difference which is chosen.

\begin{figure}
\includegraphics[width=8cm]{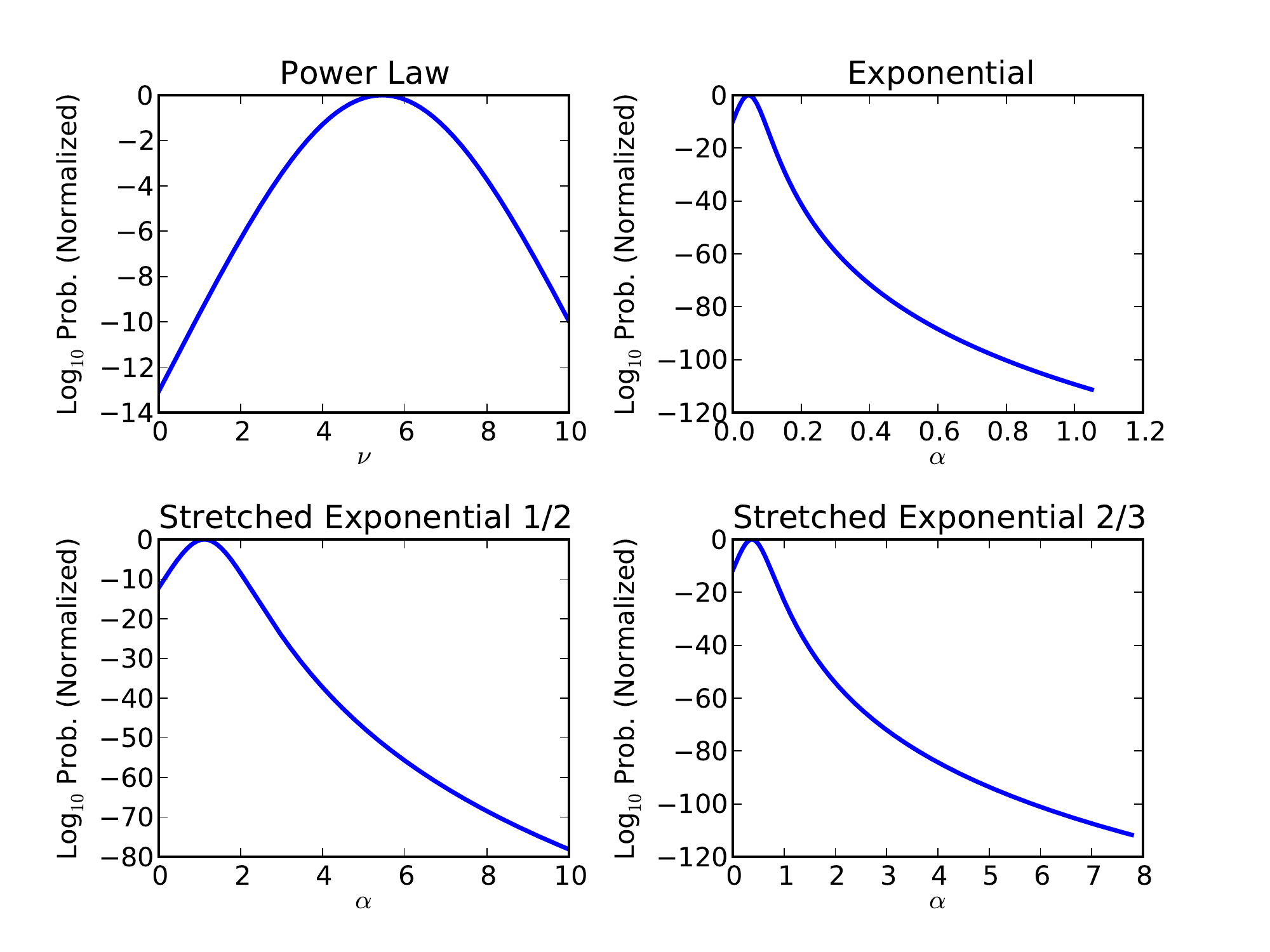}
\caption{The parameter posteriors for each model are sharply peaked around their modes - plotted here is the maximum (not the marginal value) of the posterior at various values of the key parameters $\alpha$ or $\nu$. One can see that the likelihood has a sharply peaked value along this dimension. \label{posteriors}}
\end{figure}

We have little \emph{a priori} information about the parameters we are to fit besides the fact that each ($\alpha$, $\nu$, $\sigma$) must be greater than zero. Thus, we choose a uniform prior for each over the interval $[0, \infty]$, and write this distribution $\pi( \theta)$, where $\theta$ will stand in for the vector of parameters relevant to a particular model. Then we can write the Bayesian \emph{posterior} as
\[
P( \theta | D ) \propto \mathcal{L}( D | \theta ) \pi( \theta)
\]
with $\mathcal{L}$ the likelihood from the main text. Since $\pi$ is uniform, however, we can write
\[
P( \theta | D ) \propto \mathcal{L}( D | \theta )
\]
as long as all the parameters are in the interval $[0, \infty]$. Given the posterior $P( \theta | D )$, it is common to simply take the mode as the best representative set of parameters - which are the maximum likelihood parameters used in the main manuscript. This choice is justified because we have found the posterior to be strongly peaked around the mode, as seen in Figure \ref{posteriors}.
\begin{table} 
\caption{Bayes Factors Comparing Datasets \label{bfs}}
\begin{ruledtabular}
\begin{tabular}{l c c c c}
		&		Pr. Law	& Exp.	& S. E. 1/2	& S. E. 2/3 \\
Power Law    &	 	   			 	& $1.32 \cdot 10^{5}$ 	& $3.55 \cdot 10^{1}$ 	& $4.74 \cdot 10^{2}$ \\
Exponential   & $7.55 \cdot 10^{-6}$	&	 	 				& $2.68 \cdot 10^{-4}$ 	& $4.74 \cdot 10^{2}$ \\
S. E. 1/2 	 & $2.81 \cdot 10^{-2}$ 	& $3.73 \cdot 10^{3}$ 	& 	 	 				& $1.33 \cdot 10^{1}$ \\
S. E. 2/3 	 & $2.11 \cdot 10^{-3}$ 	& $2.80 \cdot 10^{2}$ 	& $7.50 \cdot 10^{-2}$ 	& 	
\end{tabular}
\end{ruledtabular}
\end{table}

The only other difference between a Bayesian analysis and the likelihood methods are the way models are compared. In contract to a likelihood ratio, Bayesian statistics recommends a Bayes' factor $\mathcal{F}$ \cite{KASS:1995vb}, which compares two models ($M_1$, $M_2$)
\[
\mathcal{F}_{12} =  \frac{ \int_\theta \mathcal{L}_{M_1}( D | \theta ) \pi_{M_1}( \theta) }{\int_\theta \mathcal{L}_{M_2}( D | \theta ) \pi_{M_2}( \theta) }
\]
which explicitly includes information from all possible values of the parameters. The likelihood ratios used in the main text are a saddle approximation to the integrals, and thus these two methods match in the case where the posteriors are highly peaked. Table \ref{bfs} shows the calculated Bayes' factors for each model. These Bayes' factors were calculated by also integrating over $\tau_0$, the parameter accounting for a constant offset (or units) in our fits. The domain of integration for $\tau_0$ was restricted to $[-50, 10]$ for numerical stability - increasing it beyond this range resulted in little difference.

\section{Details of the Gaussian Fit Model}

Here we go over some of the details of the Gaussian model that was used to predict the folding time distribution. In the main text, we presented the central equation
\[
\int_{f^{-1}(\tau_c / \tau_0)}^{\infty} \frac{1}{\sqrt{ 2 \pi \sigma_N^2} } \exp \left[ \frac{ (N - \mu_N) ^2 }{ 2 \sigma_N^2} \right] = C
\]
from which all of our subsequent analysis begins. Using this model, we can predict $\tau_c$ for a chosen model and chosen $C$. To do this, $\mu_N$ was fixed at $105$, the empirical mode of the distribution of domain sizes (Fig.~\ref{size-dist}) -- this appeared to make subsequent fits more robust. The parameter $\sigma_N$ was then fit, via likelihood maximization, to the same empirical distribution. Subsequently, a series of values of $C$ were chosen, and these implied values of $\tau_c$ for each model, resulting in Table \ref{tau-cs}.
\begin{table} 
\caption{Most Likely $\tau_c$ for Values of $C$ (in seconds) \label{tau-cs}}
\begin{ruledtabular}
\begin{tabular}{l c c c c}
$C$		&		0.1 & 	0.05 &	0.01 &	0.001 \\
\hline
Power Law		&$1.10 \cdot 10^{1}$	&$2.88 \cdot 10^{1}$	&$1.23 \cdot 10^{2}$	&$4.36 \cdot 10^{2}$ \\
Exponential		&$3.03 \cdot 10^{2}$	&$5.85 \cdot 10^{3}$	&$1.51 \cdot 10^{6}$	&$7.64 \cdot 10^{8}$ \\
S. E. 1/2			&$9.99 \cdot 10^{1}$	&$6.54 \cdot 10^{2}$	&$1.53 \cdot 10^{4}$	&$3.47 \cdot 10^{5}$ \\
S. E. 2/3			&$1.68 \cdot 10^{2}$	&$1.57 \cdot 10^{3}$	&$7.67 \cdot 10^{4}$	&$4.25 \cdot 10^{6}$ 
\end{tabular}
\end{ruledtabular}
\end{table}

It is also possible to calculate the likelihood of observing the known empirical data given a specific $\tau_c$, via likelihood maximization. The procedure here was straightforward; first, $C$ was set to $0.10$, $\mu_N$ was again fixed at $105$, $\tau_c$ was fixed at an arbitrary value for a chosen model, and $\sigma_N$ was varied to maximize the likelihood of witnessing the empirical distribution given that model. The results for a scan of $\tau_c$ for each model is reported in Figure 3 in the main text.

\bibliography{papers2.bib}